\documentclass[%
nofootinbib,
superscriptaddress,
 preprint,
showpacs,preprintnumbers,
 amsmath,amssymb,
 aps,
pra,
 longbibliography,
 floatfix,
 lengthcheck%
]{revtex4-1}

\newcommand{\SOMeqref}[1]{\eqref{#1}}

\newcommand{\appref}[2][]{
\def\temp{#2}\ifx\temp\empty{\leavevmode\unskip Supplemental material
}\ignorespaces%
\else%
Appendix~\ifthenelse{\equal{#1}{}}{}{\ref{#1} and }%
\ref{#2}
\fi%
}

\newcommand{\MTref}[1]{\ref{#1}}
\newcommand{\MTeqref}[1]{\eqref{#1}}

\usepackage{OC-2017}
\hypersetup{breaklinks=false}

\begin{document}



\title{Structure of optimal policies in quantum control} 

\author{Dmitry V. Zhdanov}
\email{dm.zhdanov@gmail.com}
\affiliation{Northwestern Univ., Dept. Chem., Evanston, IL 60208 USA}
\author{Tamar Seideman}
\email{t-seideman@northwestern.edu}
\affiliation{Northwestern Univ., Dept. Chem., Evanston, IL 60208 USA}


\begin{abstract}
Using the Pontryagin maximum principle, the generic structure of optimal policies is deduced for typical quantum control tasks involving coherent lasers, magnetic fields and reservoir engineering. In addition, the periodic optimization is considered for the first time in view of prospective applications. We proved that nearly all optimal policies are actively constrained by technical bounds on control parameter but reduce to entirely bang-bang sequences only in special cases, such as the environmental control by random collisions. The results allow to arguably refute two generally accepted and concurring conjectures regarding the structure of optimal controls.
\end{abstract}
\maketitle
\section{Introduction\label{@SEC:Intro}}
Modern methods of NMR spectroscopy and laser coherent control (CC) allow to tackle complex practical tasks involving probing and manipulating the quantum dynamics of spins, quantum dots, atoms and molecules \cite{BOOK-Butkovskiy,BOOK-Bonnard,BOOK-Levitt,BOOK-Tannor,BOOK-d'Alessandro,2006-Shapiro}. Examples range from quantum information to chemistry and medicine. The emergent methods of quantum reservoir engineering (QRE) might further augment this list via scalability improvements and cost reductions \cite{2012-Muller,2006-Romano,2013-Kastoryano,2015-Metelmann,2013-Lemeshko,2016-Zhdanov}. But how to utilize all these technological advances most efficiently? This essentially engineering question can be best addressed using powerful methods of modern (geometric) optimal control theory developed by celebrated cohort of XX century mathematicians including McShane, Bellman, Gamkrelidze, Pontryagin and many others \cite{2014-Brockett,BOOK-Kim-2004,BOOK-Agrachev-2004,BOOK-Bonnard-2003,BOOK-Pontryagin-1962,1978-Reid}. For example, the in-depth geometric analysis of optimal quantum control of two-level systems and few special more complex cases can be found in a series of works by Boscain \cite{BOOK-Boscain,2002-Boscain,2005-Boscain,2006-Boscain,2014-Boscain}, D'Alessandro \cite{2015-Albertini,2016-Romano}, Bonnard, Sugny (with supporting experiments by Glaser group, see \cite{2012-Bonnard,2013-Garon,2014-Van_Damme,2017-Van_Damme,2017-Van_Reeth} and references therein) and others \cite{2012-Salamon,2015-Zhdanov} using Pontryagin maximum principle (PMP) \cite{BOOK-Pontryagin-1962}.

At the same time, the generic theoretical analysis of optimal policies becomes conceptually challenging already for 4-level systems \cite{2017-Duval}. Due to this fact and for historical reasons, such analysis is currently the subject of controversial speculations. A bright illustration is the theory of quantum control landscapes (TQCL) \cite{2004-Rabitz,2011-Moore,2012-Rabitz,2017-Russell} which prompted widespread beliefs that a generic globally optimal policies are easily identifiable and nearly independent on control constraints. These beliefs were criticized in a number of works \cite{2011-Pechen,2013-Fouquieres} (see Ref.~\cite{2017-Zhdanov} for recent critical review) and coexist with the opposite extreme viewpoint (see, e.g., Ref.~\cite{2017-Yang}) that the typical optimal policies are rather a bang-bang sequences where all the controls switch only between their minimal and maximal admissible values.

\begin{subequations}\label{01.01-Problem_settling}
The aim of this work is to make a step forward in resolving the above controversies for the simplest practically valuable optimization setting where the system density matrix $\hat\rho$ satisfies Markovian master equation
\begin{gather}
\tpder{}{t}\rho(t){=}\Lvn(u_1(t))\rho(t) \label{01.01.2-Liouville_equation}
\end{gather}
and the quantum Liouvillian $\Lvn$ linearly depends on a single control $u_1(t){\in}{\cal U}$:
\begin{equation}\label{01.01.3-Liouvillian_with_single_control}
\Lvn(u_1(t)){=}\Lvn_0{+}u_1(t)\Lvn_1.
\end{equation}
We additionally assume that $\Lvn_0$ and $\Lvn_1$ are some linear superoperators and that the control domain ${\cal U}$ is defined by inequalities 
\begin{equation}\label{01.01.4-Control_domains_explicit}
u_{\idx{min},1}{\leq}u_1{\leq}u_{\idx{max},1}.
\end{equation}
Depending on physical meaning of $u_1$ and $\Lvn_1$, the last term in \eqref{01.01.3-Liouvillian_with_single_control} can describe both dipole interaction with external electromagnetic field and incoherent coupling to Markovian bath. Thus, the model \eqref{01.01.2-Liouville_equation} embraces simple QRE scenarios as well as NMR and laser CC experiments governed by a single magnetic field component or linearly polarized laser field. In either case, the inequalities \eqref{01.01.4-Control_domains_explicit} represent the natural technical constraints on maximal allowed field strength or physically admissible system-bath coupling.

Assuming that $\hat O$ is a certain observable of interest, let us introduce the performance index $J{=}\Tr[{\hat O}{\rho(\tf)}]$ and consider two Mayer extremal problems
\begin{gather}\label{01.01.5-Performance_index}
J{\to}\max_{u_1{\in}{\cal U}}:
\end{gather}
\begin{description}
\item[Terminal control
] \label{03.01-DEF:terminal_control} Maximize $J$ at fixed or free final time $\tf$ starting from the given initial state $\hat\rho(\ti){=}\hat\rho_0$ at time $\ti$.
\item[Periodic control
] \label{03.01-DEF:periodic_control} 
Maximize $J(n\Tctrl)$ ($n{=}0,\pm1,\pm2...$) for asymptotic quasistationary state $\rho(t{+}n\Tctrl){=}\rho(t)$ generated by infinite periodic driving $u_1(t{+}n\Tctrl){=}u_1(t)$.
\end{description}
\end{subequations}
The second, periodic setting is rather exotic compared to the first, terminal one. Nevertheless, the ongoing developments (see, e.g., Refs.~\cite{2008-Chen,2017-Cheng,2017-Maram,2012-Xiang,2012-Tainta}) might change the situation in near future. It is worth noting that the periodic control is actively used in chemical engineering, e.g., to optimize the operation of continuous stirred tank reactors. A thorough review can be found in Ref.~\cite{BOOK-Silveston}.

The rest of the letter contains detailed generic analysis of the problem \eqref{01.01-Problem_settling}. After formulating our central theorem we will discuss its physical meaning with emphasis on periodic, CC and QRE cases. We will conclude with a brief summary and outlook.



\section{The main result\label{@SEC:Math}} We will approach the problem \eqref{01.01-Problem_settling} using the framework of Pontryagin maximum principle (PMP) \cite{BOOK-Kim-2004,
BOOK-Milyutin-1998} which is briefly reviewed in \appref{@SEC:APP-PMP}. PMP supplies the first-order necessary optimality conditions for virtually all kinds of control constraints, including the case of Eq.~\eqref{01.01.4-Control_domains_explicit}. PMP exploits the properties of Pontryagin's function (PF) $K(\rho(t),u_1(t),\psi(t),t)$ of four arguments, where $\psi(t)$ is the matrix of costate variables satisfying the equations $\pder{\psi_{i,j}}{t}{=}{-}\pder{K}{\rho_{i,j}}$ and special boundary transversality conditions. Specifically, an optimal control $\tilde u_1$ maximizes $\tilde K(u_1(t),t){\equiv}K(\tilde\rho(t),u_1(t),\tilde\psi(t),t){=}\const$ along the optimal state $\tilde\rho$ and costate $\tilde\psi$ trajectories: $\tilde u_1(t){=}\arg\max_{u_1(t)}\tilde K(u_1(t),t)$. One can distinguish two ways in which the segments (subarcs) of optimal trajectory $\tilde u_1(t)$ can obey PMP: (i) if
$\pder{\tilde K(\tilde u_1)}{u_1(t)}\genfrac{}{}{0pt}{}{{>}}{({<})}0$, then the subarc is pinned to the boundary: $\tilde u_1(t){=}u_{\idx{max(min)},1}$, and called \emph{regular}; (ii) otherwise the subarc is \emph{singular} and ``walks'' somewhere inside the domain $\cal U$. 

It is worth noting that there exist some clashes between the above generally accepted PMP terminology and the TQCL-specific language. To avoid confusions, the distinctions are summarized in \appref{@SEC:APP-notations&pitfalls}. 

The PF of problem \eqref{01.01-Problem_settling} linearly depends on $u_1(t)$: 
\begin{equation}\label{02.01-Pontryagin's_function}
K(\rho(t),u_1(t),\psi(t),t){=}\Tr[{\psi(t)}{\Lvn(u_1(t))}{\rho(t)}].
\end{equation}
The extremal $\{\tilde\rho(t),\tilde\psi(t),\tilde u_1(t)\}$ may be uniquely defined by the state and costate variables ${\tilde\psi(\ti)}$, $\tilde\rho(\ti)$ at initial time $t{=}\ti$ and certain additional parameters $p_k$ characterizing the endpoint $t{=}\tf$ and the junction points between subarcs where either $\tilde u_1(t)$ 
or any of its time derivatives has discontinuity (see \appref{@SEC:APP-extremals} for details). 
%

The extremal problem \eqref{01.01-Problem_settling}, thus, can be equivalently restated as a problem of finding $P_{\idx{total}}$ unknown continuous parameters $\rho(\ti)$, ${\psi(\ti)}$ and  $p_k$ satisfying certain inequalities and $C_{\idx{total}}$ equations of equality type. These $C_{\idx{total}}$ equations include the boundary constraints (known as \emph{transversality conditions}) and special restrictions (the generalizations of Weierstrass-Erdmann corner conditions) that must be obeyed at junctions. The values of $P_{\idx{total}}$ and $C_{\idx{total}}$ depend on the number of junctions and the ordering of regular and singular subarcs. In the cases when all $C_{\idx{total}}$ constraints are independent the solutions $\tilde u_1$ of optimal problem \eqref{01.01-Problem_settling} should almost always obey the inequality $P_{\idx{total}}{\geq}C_{\idx{total}}$. The following generic theorem is the central result of this letter.
\begin{theorem}[See \appref{@SEC:APP-lemma} for proof]\label{01.01-Lemma-1_substitute}
The control policies $u_1(t)$ for which $P_{\idx{total}}{\geq}C_{\idx{total}}$ must start and end with regular arcs. Also, all (if any) their continuous singular segments must be $C^{\infty}$-smooth. 
\end{theorem}
\noindent
Theorem \ref{01.01-Lemma-1_substitute} suggests that the solution $\tilde u_1$ of a typical problem \eqref{01.01-Problem_settling} approaches the boundary of ${\cal U}$ regardless of choice of $u_{\idx{min},1}$ and $u_{\idx{max},1}$ in Eq.~\eqref{01.01.4-Control_domains_explicit}. However, care is needed regarding the required independence of $C_{\idx{total}}$ constraints when applying this result. The matter is that the constraints redundancies are inherent to several physically important cases. It is worth to consider these exceptions to clarify the physical meaning of the generic result.

\subsection{Exception 1: Terminal control of closed system}
A common starting point to model certain CC and NMR experiments is to assume that the quantum system is completely isolated from the dissipative environment. In this case, the terms in rhs of Eq.~\ref{01.01.3-Liouvillian_with_single_control} take the form 
\begin{gather}\label{02.01-closed_dipole_case}
\Lvn_0{=}{-}\frac{i}{h}[\hat H_0,\rho],~~\Lvn_1{=}\frac{i}{h}[\hat\mu,\rho],
\end{gather}
where 
$\hat H_0$ is eigen Hamiltonian and $\hat\mu$ is the operator of electric or magnetic dipole momentum component along the direction of control field $u_1(t)$. The corresponding formal solution of Eq.~\eqref{01.01.2-Liouville_equation} is a unitary transformation $\rho(t_2){=}\hat U_{t_2,t_1}(u_1){\rho(t_1)}\hat U^{-1}_{t_2,t_1}(u_1)$, where $\hat U_{t_2,t_1}(u_1){=}\overrightarrow{\exp}\left\{{-}\int_{t_1}^{t_2}\frac ih(\hat H_0{-}u_1(t)\hat\mu_x)dt\right\}$. The criterion $\pder{\tilde K(\tilde u_1)}{u_1}\big|_{t{=}t'}{=}0$ of the singular subarc at $t{=}t'$ can be restated in terms of $\hat U$ as
\begin{gather}\label{02.01-main_lemma_equation}
\Tr\left[\left[\tilde\rho(t_0),\tilde O(t_0)\right]\hat U_{t',t_0}^{-1}(\tilde u_1)\hat \mu \hat U_{t',t_0}(\tilde u_1)\right]{=}0,
\end{gather}
where $\tilde{O}(t){=}\hat U_{\tf,t}^{-1}(\tilde u_1)\hat{O}\hat U_{\tf,t}(\tilde u_1)$ and $t_0{\in}[\ti,\tf]$ is an arbitrary time instant.

It has been long known that the problem \eqref{01.01-Problem_settling} with Liouvillian \eqref{02.01-closed_dipole_case} allows for globally optimal solutions satisfying $[\tilde \rho(t),\tilde O(t)]{=}0$ provided that rather mild Lie algebra rank condition (see. e.g., \cite{BOOK-d'Alessandro}, Sec.~3.2) on $\hat H_0$ and $\hat\mu$ is satisfied
and the control time $\Tctrl{=}\tf{-}\ti$ is long enough: $\Tctrl{>}\Topt$ \cite{1972-Jurdjevic}. Note that the condition $[\tilde \rho(t),\tilde O(t)]{=}0$\footnote{This case is referred as ``kinematic critical point'' in the TQCL (see \appref{@SEC:APP-notations&pitfalls} for details).} implies that the criterion \eqref{02.01-main_lemma_equation} is automatically fulfilled everywhere on the extremal. Such situations are called degenerate and obviously violate the assumptions of theorem \ref{01.01-Lemma-1_substitute}. In particular, the optimal controls $\tilde u_1(t)$ in this case can be essentially nonanalytic and contain any number of discontinuities. 

In contrast, if control time is not sufficiently long, $\Tctrl{<}\Topt$, then $[\tilde\rho(t),\tilde O(t)]{\ne}0$, and the $\tilde u_1(t)$ is expected to match the predictions of theorem~\ref{01.01-Lemma-1_substitute} in absence of other occasional constraints redundancies. The results of in-depth studies of the controlled two-level system  (see, e.g., \cite{2012-Pechen,2015-Zhdanov,BOOK-Agrachev-2004,BOOK-Boscain,BOOK-d'Alessandro}) can serve an excellent and simple illustration of this transition.


\subsection{Exception 2: Terminal control of thermalized open system} Perfectly closed quantum system is unrealistic idealization. In practice, any system Liouvillian $\Lvn_0$ includes some dissipative terms responsible for system-bath interactions. These terms are pledge of inevitable equilibration into thermal state $\rhotd$, such that $\Lvn_0\rhotd{=}0$. Here we will consider the practically important situation when $\hat\rho(\ti){=}\rhotd$. As in the case of closed system, denote as $\Topt$ the minimal control time at which the maximal possible value of $J$ can be achieved. It may be shown that PF of optimal policy $\tilde u_1$ is constant and positive (zero) along the trajectory for $\Tctrl{<}\Topt$ $(\Tctrl{\geq}\Topt)$. In the case $\Tctrl{<}\Topt$ we have: $\tilde K|_{t=\ti}{=}\Tr[{\psi(\ti)}{\Lvn(u_1(\ti))}{\rhotd}]{=}\tilde u_1(0)\Tr[{\tilde\psi(\ti)}{\Lvn_1}{\rhotd}]{>0}$. Hence, according to PMP, the optimal control $\tilde u_1(\ti)$ reaches the boundary of ${\cal U}$ at $t{=}\ti$, that is, the first arc is regular in agreement with theorem \ref{01.01-Lemma-1_substitute}.

However, in the case $T{\geq}\Topt$ the transversality condition $K{=}0$ implies that $\pder{\tilde K(\tilde\vecu)}{u_1}\big|_{t{=}\ti}{=}\ti$, thus, effectively reducing the number of independent boundary constraints and making the optimal solutions with singular first arc a viable possibility. Furthermore, the conclusions of theorem \ref{01.01-Lemma-1_substitute} remain applicable to the rest of optimal extremal. In particular, unless we have some additional occasional redundancies, the second ark is expected to be regular one. The physical meaning of a possible optimal solution of this sort is quite obvious: One leaves system uncontrolled (keeping $u_1{=}0$) for $t{-}\ti{<}\Tctrl{-}\Topt$ and then applies the optimal solution for $\Tctrl{=}\Topt$ during the remaining time. 
In other words, PMP naturally confirms the intuitively evident advantages of controlling dissipative systems at the shortest possible time, which stimulate increasing interest to quantum speed limit problems and shortcuts to adiabaticity (see, e.g., \cite{2017-Campbell} for references). 

Just discussed example hints the physical interpretation of the generic result of theorem \ref{01.01-Lemma-1_substitute}. Namely, the dissipative part of $\Lvn_0$ leads to usually unrecoverable and undesirable modifications of any initial system state $\hat\rho(\ti){\ne}\rhotd$. To mitigate this effect, it is critical to quickly transfer the initial state $\hat\rho(\ti)$ into the suitable subspace maximally protected against dissipation (relative to the control objective). This goal gives rise to the starting regular arc. Similarly, the optimal final state generally lies outside the dissipation-prone subspace. Hence, the final state preparation step also should be done quickly. The latter explains why the ending arc of $\tilde u_1(t)$ is expected to be regular too.

It is worth noting that the above analysis contravenes the central statement of TQCL that under reasonable physical assumptions the control bounds, Eq.~\eqref{01.01.4-Control_domains_explicit}, are irrelevant once they are relaxed enough. However, what if the physical nature of control parameter allows to treat it as unconstrained? Intuitively, one would expect all the regular arcs to collapse into delta-function-like bumps. However, this implies that $\Topt{\to}0$ for purely bang-bang optimal solutions in this limit. At the same time, the impossibility of instant control is apparent for majority of practical cases. The contradiction disappears if one admits that the typical optimal solutions consist of combinations of both regular and singular segments. The following analysis of periodic control will numerically confirm the conclusions of this informal reasoning. Moreover, we will see that a finite $\Topt$, and hence the globally optimal solution, do not exist at all for this kind of optimization. Besides, we will also unravel one more situation where the 
assumptions of theorem \ref{01.01-Lemma-1_substitute} are violated.


\section{Periodic control\label{@SEC:Control-periodic}}
The quantum optimal periodic control offers opportunity to asymptotically stir the \emph{arbitrary} initial state $\rho({-}\infty)$ of open quantum system into the same quasistationary final state $\tilde\rho(t)$ reviving with period $\Tctrl$. Despite the periodic driving is important method of coherent control \cite{2017-Eckardt}, the theory of periodic optimization for last 50 years was primarily developed in chemical engineering context \cite{BOOK-Silveston} where the problem dimensionalities usually are not too high but the dynamical equations are highly complex and nonlinear. This work seems to be the first attempt to extend the theory to the case of controllable quantum dynamics of form \eqref{01.01.2-Liouville_equation}.

The following statement can be proven (see \appref{@SEC:APP-periodic_control_theorem}):
\begin{theorem}\label{05.01-periodic_control_theorem}
The solution $\tilde u_1(t\mathrm{~mod~}\Tctrl)$ of the periodic control problem \eqref{01.01-Problem_settling} can be globally optimal iff it is also extremal for the respective terminal problems with the control times $n\Tctrl$, for any integer $n{=}2,3,...$ and the initial state $\tilde\rho(0)$.
\end{theorem}

\noindent To our knowledge, there is no evidence of the asymptotically time-periodic solutions $\tilde u(t)$ for quantum terminal problems. Thus, theorem~\ref{05.01-periodic_control_theorem} indicates that every extremal $\tilde u_1$ of a typical periodic control problem \ref{01.01-Problem_settling} is only locally optimal (the ``trap'' in TQCL terminology) and can be improved via doubling the period $\Tctrl$. 

\begin{figure}[t]
\includegraphics[width=0.45\textwidth]
{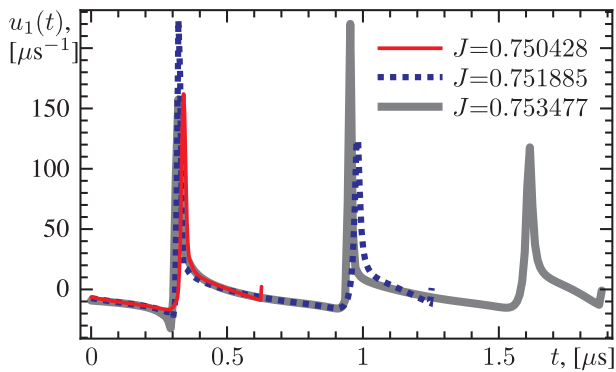}
\caption{Three different locally optimal controls $\tilde 
u_1(t~\rm{mod}~T)$ for the periodic control of coherence in the $\Lambda$-system (one period is shown). The detailed specification of the model and its parameters can be found in \protect\appref{@SEC:APP-Lambda-system}.\label{@FIG.01}}
\end{figure}

As an example, consider the nondegenerate open $\Lambda$-system $1{\LR}3{\LR}2$ subjected into three radiative decay channels $3{\leadsto}2$, $3{\leadsto}1$, $2{\leadsto}1$ and the laser field $\vec{\cal E}{=}\vec A\cos(\omega t{+}\int_{{-}\infty}^{t}u_1(t)dt)$ with constant flux 
and periodically varying controlled instant frequency $\omega{+}u_1(t)$. Note that it might be legitimate to treat the control $u_1$ as unbounded if the laser spectrum is broad enough.
The Liouvillian of this system within the rotating wave approximation takes the form \eqref{01.01.3-Liouvillian_with_single_control}. We numerically solved the extremal problem of periodic $1{\LR}2$-coherence enhancement: $\rho_{12}{+}\rho_{21}{\to}\max$, in absence of explicit constraints on $\cal U$ and with free period $\Tctrl$. The results are shown in Fig.~\ref{@FIG.01}. 
One can see that the extremals corresponding to larger locally optimal periods $\Tctrl$ deliver larger values to performance index $J$ in agreement with theorem \ref{05.01-periodic_control_theorem}. The sharp pikes observed near $0.4$, $1.0$, and $1.6~\mu$s are the limiting cases of regular arcs for unbounded control. They are found to approximate the scaled delta functions $\pi\delta(t)$. Thus, these are the time instants of sudden reverse of field direction. 

An additional small sharp pikes can be also noticed for each policy at endpoints $t{=}\Tctrl$. Nevertheless all the calculated extremals begin with a singular arc. According to theorem \ref{01.01-Lemma-1_substitute}, this is a sign of certain redundancy among the boundary constraints. Indeed the chosen observable $\hat O$ in the performance index \eqref{01.01.5-Performance_index} commutes with dipole momentum operator $\hat\mu$. The latter leads to the equality $\hat O\Lvn_1{=}0$ which introduces redundancy into the periodic transversality conditions (see \appref{}, Eq.~\SOMeqref{APP_PMP-transversality-conditions}).

The presented findings clearly controvert the conjecture made in Ref.~\cite{2017-Yang}  that the singular arcs are non-generic for optimal solutions of quantum problems. However, are there special cases in which this conjecture indeed holds? We will conclude our analysis by showing that one such case is QRE, an emerging technology to control quantum dynamics using non-conservative forces and dissipative processes.


\section{Control via quantum reservoir engineering (QRE)\label{@SEC:Control-environment}}
Here we consider the generalized variant of problem \eqref{01.01-Problem_settling} with $\Nctrl$ independent linear controls $\vecu{=}\{u_1,...,u_{\Nctrl}\}$ and quantum Liouvillian (cf. Eq.~\eqref{01.01.3-Liouvillian_with_single_control})
\begin{gather}\tag{1b*}
\Lvn(u_1(t)){=}\Lvn_0{+}\textstyle\sum_{k{=}1}^{\Nctrl}u_k(t)\Lvn_k. \label{01.01.3-Liouvillian}
\end{gather}
The range of each control $u_k$ is assumed to be constrained similarly to Eq.~\eqref{01.01.4-Control_domains_explicit}. We are interested in the special case when some (or all) of the superoperators $\Lvn_k$ in \eqref{01.01.3-Liouvillian} represent the dissipative couplings to an independently controllable reservoirs and have form
\begin{gather}\label{04.01-Liouvillian_in_model_of_random_collisions}
\Lvn
_k\rho{=}{-}\rho{+}\rho_{k},
\end{gather}
where $\rho_{k}$ describes the effective equilibrium state relative to $k$-th dissipative channel. The supeoperators of form \eqref{04.01-Liouvillian_in_model_of_random_collisions} can capture the effects of strong inelastic random collisions with tunable rates $u_k$ 
 and also well-approximate many other dissipative mechanisms when the deviations from the equilibrium $\rho_{\mathrm{td}}$ are small \cite{BOOK-Ilinskii-1994}. 

The following theorems (see \appref[@SEC:APP-projective_control_theorem]{@SEC:APP-mixed_control_theorem} for their proofs) show that the corresponding generalized optimal control problem \ref{01.01-Problem_settling} shares many characteristic features with the celebrated linear time-optimal problems (\emph{cf. }\cite{BOOK-Pontryagin-1962}, p.120 or \cite{BOOK-Agrachev-2004}, Ch.~15).

\begin{theorem}\label{04.01-projective_control_theorem}
Suppose that all the controlled parts $\Lvn_k$, $k{=}1,...,\Nctrl$ of the Liouvillian \eqref{01.01.3-Liouvillian} are of form \eqref{04.01-Liouvillian_in_model_of_random_collisions}. Then, the optimal solution $\tilde\vecu$ of the generalized problem \eqref{01.01-Problem_settling} is a piecewise-constant bang-bang control.
\end{theorem}

The next theorem expands the above result to the case of Liouvillian with controls of the mixed type.
\begin{theorem}\label{04.01-mixed_control_theorem}
Suppose that the Liouvillian \eqref{01.01.3-Liouvillian} of the generalized terminal control problem \eqref{01.01-Problem_settling} has the term $\Lvn_{k}$ of form \eqref{04.01-Liouvillian_in_model_of_random_collisions}. Then, the associated optimal control $\tilde u_k(t)$ cannot comprise singular subarcs, except for the cases when $\tilde u_k$ is unspecified and redundant.
\end{theorem}

Interestingly, the bang-bang optimal dissipative control policies are de facto ubiquitously incorporated into CC experiments for decades. Indeed, a typical experiment begins with steering the system into suitable initial state (usually the ground one) via contact with appropriate thermostat, whereas the subsequent manipulations involving coherent radiation are normally performed under conditions of maximal isolation. Theorem \ref{04.01-mixed_control_theorem} delivers a formal rationale to this common scenario.



\section{Discussion and outlook\label{@SEC:Discussion}} We have demonstrated that our primary result, theorem \ref{01.01-Lemma-1_substitute} encapsulates valuable generic information about quantum optimal policies for terminal and periodic quantum control of complex quantum systems and allows to treat a broad variety of interactions, from fully coherent to strictly dissipative. Nevertheless, several practically important issues remain out of scope of presented analysis. We did not address the technical peculiarity of laser-driven optimal control where the constraints on control fields are imposed in frequency rather than in time domain. We also did not account for the fact that the experimentally attainable best policies are pareto-optimal and compromise high efficacy with low sensitivity to the inevitable fluctuations and uncertainties of inputs and controls
\footnote{For instance, in our numerical example on periodic control of the $\Lambda$-system (Fig.~\ref{@FIG.01}), a simple harmonic control $u_1(t){=}{-}1.58{+}1.61\cos(\frac{2\pi t[\mu{\mathrm s}]}{0.626})$ [MHz] earns the performance index which is just 7\% less compared to the formal optimal solution. At the same time, there exist established techniques for producing stable and robust harmonic frequency modulation \cite{2000-Hall}. For this reason, this simpler control policy might actually outperform in laboratory.}. Lastly, our analysis is limited to Markovian dynamics. However, the results of recent study \cite{2016-Romano} of a simple non-Markovian case also match the generic predictions of theorem \ref{01.01-Lemma-1_substitute}. Additional research is needed to unravel systematic or occasional origin 
 of this coincidence.

Despite all these limitations, our analysis allows to arguably reject as a primitive oversimplifications both the conjecture of TQCL on practical irrelevance of control constraints under reasonable physical assumptions as well as the opposite viewpoint of Ref.~\cite{2017-Yang} about expected vast predominance of bang-bang policies%
\footnote{The authors of \cite{2017-Yang} conjecture that this is the case for complex disordered quantum systems even though the numerical results shown in Fig.~2 of their work seem to disagree with the authors' claim. Furthermore, the work \cite{2017-Yang} does not provide any justifications (such as an evidence of constancy of the Pontryagin function along the trajectory, its correct behavior at corner points) that the identified bang-bang policies indeed satisfy PMP.}. In particular, we have shown that entirely singular optimal policies are essentially the numerical artifacts of the closed system approximation. Such policies will be dramatically restructured by dissipation in laboratory optimization. At the same time, the bang-bang policies can indeed be optimal in certain cases including incoherent control by collisions. 


\bibliographystyle{apsrev4-1}
\nocite{apsrev41Control}
\bibliography{OC-2017_bibstyle,OC-2017}

\begin{thebibliography}{57}
\bibitem [{REF()}]{REFINEMENT-notations}%
  \BibitemOpen
  \href@noop {} {}\bibinfo {note} {Hereafter we will everywhere use the bra-ket
  notations introduced in Appendix~\ref{@SEC:APP-bra-ket}.}\BibitemShut {Stop}%
\end{thebibliography}
\onecolumngrid
\bibsection
\begin{center}
\Large\protect{\texttt{\uppercase{Supplemental material}}}
\end{center}
\twocolumngrid
\appendix
\setcounter{section}{0}

\setlength\heavyrulewidth{0.10em} 
\setlength\lightrulewidth{0.06em} 
\newcommand{\SOMTITLE}{supplemental material}

\section*{Introduction}
In this \SOMTITLE, we briefly review the basics of the theory of Pontryagin maximum principle (Appendix~\ref{@SEC:APP-PMP}) emphasizing the case of linear control. Then, in Appendix~\ref{@SEC:APP-bra-ket} we restate the optimal quantum control problem employing the convenient Dirac formalism which will be used in the rest of this \SOMTITLE. Appendix~\ref{@SEC:APP-notations&pitfalls} is aimed to disambiguate the controversial notions and terminology used in the modern literature. The next Appendix~\ref{@SEC:APP-extremals} reviews the sewing conditions between segments of optimal policies for problems with a single control parameter. Then, Appendices~\ref{@SEC:APP-lemma} -- \ref{@SEC:APP-periodic_control_theorem} present correspondingly the detailed proofs of four statements made in the main text. Finally, Appendix~\ref{@SEC:APP-Lambda-system} provides the detailed specification of the controlled $\Lambda$-system used in our numerical example. 

\section{Review of Pontryagin maximum principle\label{@SEC:APP-PMP}}
In this appendix we briefly summarize the key statements of Pontryagin theory for completeness and integrity of the presentation (for more details see e.g.~\cite{BOOK-Kim-2004}). In its canonical geometrical settling, this theory addresses the following endpoint problem \cite{BOOK-Agrachev-2004}:
\begin{subequations}\label{APP_PMP-pr-st}
\begin{gather}
g_0(\vecx(\tf)){=}x_0(\tf){\to}\max\label{APP_PMP-pr-st-perf-idx},\\
\pder{}{t}x_i{=}f_i(\vecx,\vecu,t)~~(i{=}0,...,n),\label{APP_PMP-pr-st-dyn-law}\\
\vecu{\in}{\cal U},\notag\\
g_j(\vecx(\ti),\vecx(\tf),\ti,\tf){=}0~~(j{=}1,...,q{<}2n{+}2),\label{APP_PMP-pr-st-bnd-cnst}
\end{gather}
\end{subequations}
where the meanings of $\vecu$ and $\cal U$ are the same as in the main text, $\vecx$ is the set of the state (or phase) variables and $g_j$ define the boundary constraints. Importantly, the set ${\cal U}$ of admissible controls here in principle can be virtually any manifold. This fact allows to apply theory, e.g., to the cases when controls $\vecu$ can take only discrete values. In practice, the phase variables $\vecx'{=}\{x_i\}$ with $i{>}0$ usually describe the dynamics of the physical system, and $f_i(\vecx,\vecu,t){\equiv}f_i(\vecx',\vecu,t)$ while the extra coordinate $x_0$ defines the \emph{performance index} $J{=}x_0(\tf)$ to be optimized. Specifically, the Bolza problem
\begin{gather}
J{=}P(\vecx'(\ti),\vecx'(\tf),\ti, \tf){+}\int_{\ti}^{\tf}Q(\vecx',\vecu,t)dt\to\max\label{APP_PMP-pr-st-perf-idx-2}
\end{gather}
can be defined by setting $x_0(\ti){=}P(\vecx'(\ti),\vecx'(\ti),\ti, \tf)$ and $f_0(\vecx',\vecu,t){=}Q(\vecx',\vecu,t){+}\sum_{i=1}^N\tpder{P(\vecx'(\ti),\vecx'(t),\ti, T)}{x_i(t)}f_i(\vecx',\vecu,t)\notag$. The particular cases when the first (second) term in  \eqref{APP_PMP-pr-st-perf-idx-2} are absent represent the Lagrange (Mayer) problems.

Let us introduce the following auxiliary functions:
\begin{align}
&K{=}\sum_{i{=}0}^N\Psi_if_i&&\mbox{--- Pontryagin function (PF);}\label{APP_PMP-PF}\\
&G{=}\sum_{j{=}0}^{q}\nu_jg_j&&\mbox{--- terminant,}\label{APP_PMP-terminant}
\end{align}
where $\nu_0,\Psi_0{=}\const{\geq}0$ and the $\Psi(t)$ stands for the set of so-called costate (or adjoint) variables. By definition,
\begin{subequations}\label{APP_PMP-dyn-eqs}
\begin{align}
&\pder{}{t}x_i{=}\pder{K}{\Psi_i} \qquad\mbox{(\emph{cf.} \eqref{APP_PMP-pr-st-dyn-law})};\\
&\pder{}{t}\Psi_i{=}{-}\pder{K}{x_i}.
\end{align}
\end{subequations}
Mathematically, the functions $\Psi(t)$ and variables $\nu$ represent the Lagrange multipliers to handle the dynamic and boundary constraints \eqref{APP_PMP-pr-st-dyn-law} and \eqref{APP_PMP-pr-st-bnd-cnst} in the extremal problem \eqref{APP_PMP-pr-st-perf-idx}. The process (trajectory) $v(t){=}\{\Psi(t),\vecu(t),\vecx(t)\}$ is called \emph{admissible} if it matches Eqs.~\eqref{APP_PMP-dyn-eqs} and boundary conditions \eqref{APP_PMP-pr-st-bnd-cnst}.

The \emph{Pontryagin maximum principle} (PMP) states that each (locally) optimal solution of problem~\eqref{APP_PMP-pr-st} (hereafter labeled with $\tilde~$) is represented by the process $\tilde v(t){=}\{\tilde\Psi(t),\tilde\vecu(t),\tilde\vecx(t)\}$ where $\tilde\Psi_0{\geq}0$ and $\tilde\Psi(t){\neq}0$, such that:
\begin{gather}\label{APP_PMP-PMP}
\tilde\vecu(t){=}\arg\max_{\vecu(t){\in}{\cal U}}K(\tilde\Psi(t),\vecu(t),\tilde\vecx(t),t),
\end{gather} 
and the following \emph{transversality conditions} hold:
\begin{subequations}\label{APP_PMP-transversality-conditions}
\begin{align}
\Psi_i(\ti)&{=}{-}\pder{G}{x_i(\ti)} & \Psi_i(\tf)&{=}\pder{G}{x_i(\tf)}\label{APP_PMP-transversality-conditions-1}\\
\left.K\right|_{t{=}\ti}&{=}\pder{G}{\ti} & \left.K\right|_{t{=}\tf}&{=}{-}\pder{G}{\tf}\label{APP_PMP-transversality-conditions-2}
\end{align}
\end{subequations}

Any processes satisfying Eqs.~\eqref{APP_PMP-PMP} and \eqref{APP_PMP-transversality-conditions} are called \emph{extremals}. The extremal is not necessarily the solution of the problem \eqref{APP_PMP-pr-st} since PMP provides only the first-order necessary optimality condition. The solutions often can be identified using the Legendre-Clebsch condition and its generalizations \cite{1977-Krener}, or other higher-order extensions of PMP~\cite{BOOK-Milyutin-1998}. 

One distinguishes \emph{regular} and \emph{singular} (or \emph{degenerate}) extremals{\textbackslash}problems. In the first case the optimal control $\tilde\vecu$ can be directly obtained from Eq.~\eqref{APP_PMP-PMP} while in the second one an extra investigation is required. The optimal trajectory may also be a combination of regular and singular subarcs, and the functions $\tilde u_i(t)$ may have any number of discontinuities of the first kind (\emph{corner points}) both in the interior of each arc as well as at their junction points. The following \emph{Weierstrass-Erdmann conditions} for costate $\Psi$ and PF must hold at each corner point $t'$:
\begin{gather}\label{APP_PMP-sewing-conditions}
\left.\Psi\right|_{t'-0}{=}\left.\Psi\right|_{t'+0};\qquad \left.K\right|_{t'-0}{=}\left.K\right|_{t'+0}
\end{gather}

The extremal problems considered in this paper belong to the special case when the PF \eqref{APP_PMP-PF} linearly depends on the controls $\vecu$. For these problems it is convenient to define the \emph{switching functions} $\tilde K_{u_i}(t){=}\pder{K(\tilde\Psi(t),\vecu(t),\tilde\vecx(t))}{u_i(t)}$. It follows from the PMP \eqref{APP_PMP-PMP} that $\tilde K_{u_i}(t){\ne}0$ is the signature of regular optimal controls of bang-bang type:
$u_i(t){=}\arg\max_{u_i(t)\in{\cal U}}\sign(\tilde K_{u_i}(t))u_i(t)$. Correspondingly, $\tilde K_{u_i}(t){=}0$ benchmarks the point on a singular subarc. The extremals containing both the regular and singular parts are often referred as the bang-singular.

Sometimes the optimal solution corresponds to $\Psi_0{=}0$. This happens when the solution does not depend on the performance index or there exists only one admissible trajectory and in other ill-posed problem settlings. The corresponding problems are called \emph{abnormal}.


\section{Statement of quantum control problem in bra-ket notations\label{@SEC:APP-bra-ket}}

This appendix introduces the reformulation the generalized optimal control problem \MTeqref{01.01-Problem_settling} with Liouvillian \MTeqref{01.01.3-Liouvillian} in convenient Dirac bra-ket notations. These notations naturally account for the linearity of Liouville equation \MTeqref{01.01.2-Liouville_equation} in the state variables $\rho$ and both allow to improve the presentation and essentially simplify the proofs of the statements made in the main text. To avoid confusion, we stress that  in these appendices the Dirac notations are \textbf{not} used for denoting the wavefunctions of pure quantum states unless it is explicitly stated.

Assume that our controlled system is $N$-dimensional. Define the $N^2$-dimensional orthogonal Hermite matrix basis $\vecsigma$ consisting of arbitrary Hermite matrices $\sigma_i{\in}\Cmpl^{N{\times}N}$ satisfying the relations: $\Tr[\sigma_i\sigma_j]{=}\delta_{i,j}$. Using this basis, any Hermitian operator $\hat O{\in}\Cmpl^{N{\times}N}$ can be bijectively expanded to the real vector $\ket{O}{\in}\Real^{N^2}$ with elements $O_i{=}\Tr[\sigma_i\hat O]$, such that $\hat O{=}\sum_iO_i\sigma_i$. Similarly, the Liouville superoperator $\Lvn$ 
can be mapped to ${N^2{\times}N^2}$-dimensional real matrix $\sLvn$ 
with elements $\sLvn_{ij}{=}\Tr[\sigma_i\Lvn\sigma_j]$. Using these notations, Eqs.~
\MTeqref{01.01.5-Performance_index}, \MTeqref{01.01.2-Liouville_equation} and  \MTeqref{01.01.3-Liouvillian} can be restated as:
\begin{subequations}\label{SOM-01.01-Problem_settling}
\begin{align}
&J{=}\scpr{O}{\rho(\tf)}{\to}\max_{\vecu{\in}{\cal U}}, \label{SOM-01.01.1-Objective}\\
&\pder{\ket{\rho}}{t}{=}\sLvn(t)\ket{\rho} \label{SOM-01.01.2-Liouville_equation}\\
&\sLvn(t){=}\sLvn_0{+}\sum_{k}u_k(t)\sLvn_k, \label{SOM-01.01.3-Liouvillian}
\end{align}
\end{subequations}
The problem \eqref{SOM-01.01-Problem_settling} can be put in the form \eqref{APP_PMP-pr-st} by introducing an extra state variable $x_0$ evolving according to dynamic law: $\pder{x_0}{t}{=}0$, 
and satisfying the boundary constraint: $x_0(\tf){=}\scpr{O}{\ket{\rho(\tf)}}$. Then, the state vector $\vecx{\in}\Real^{N^2{+}1}$ in \eqref{APP_PMP-pr-st} reads $\{x_0,\bra{\rho}\}^\transpose$. Introducing the similar expansion $\{\Psi_0,\bra{\psi}\}^\transpose$ for adjoint variables $\Psi{\in}\Real^{N^2+1}$, we are able to write the PF \eqref{APP_PMP-PF} in the form:
\begin{gather}\label{SOM-02.01-Pontryagin's_function}
K{=}\matel{\psi(t)}{\sLvn(u(t))}{\rho(t)},
\end{gather}
which is identical to Eq.~\MTeqref{02.01-Pontryagin's_function}. The corresponding sets of transversality conditions \eqref{APP_PMP-transversality-conditions-1} are specified as follows.
\begin{description}
\item[Terminal control]
\begin{gather}\label{SOM-03.01-DEF:terminal_control}
\bra{\psi(\tf)}{=}\bra{O_{\idx{n}}},~\ket{\rho(\ti)}{=}\ket{\rho_{\idx{i}}}\\
\mbox{+ normalization conditions}\nonumber.
\end{gather}
\item[Periodic control]
\begin{gather}\label{SOM-03.01-DEF:periodic_control}
\bra{\psi(\ti)}{=}\bra{\psi(\tf)}{-}\bra{O_{\idx{n}}},~
\ket{\rho(\tf)}{=}\ket{\rho(\ti)}\\
\mbox{{+}normalization conditions}\notag, 
\end{gather}
\end{description}
where $\bra{O_{\idx{n}}}{=}\bra{O}{-}\scpr{O}{\rho(\tf)}\bra{1}$, $\bra{1}$ denotes the vector representation of the identity operator, and the normalization conditions are:
\begin{gather}\label{03.01-psi_&_rho_normalization}
\scpr{\psi(t)}{\rho(t)}{=}0;~\scpr{1}{\rho(t)}{=}1.
\end{gather} 
The second set \eqref{APP_PMP-transversality-conditions-2} of transversality conditions produces an additional constraint only in the case of free control time $\Tctrl$:
\begin{gather}\label{SOM-03.01-K=0(free_time)}
K(t){=}0~~\mbox{(for non-fixed control time $\Tctrl$ only).}
\end{gather}
Here we also used the fact that PF of problem \eqref{SOM-01.01-Problem_settling} is constant along the extremal. The latter can be straightforwardly checked by considering the sequential time derivatives of Eqs.~\eqref{SOM-02.01-Pontryagin's_function} and applying PMP \eqref{APP_PMP-PMP}.

\section{Terminological conventions\label{@SEC:APP-notations&pitfalls}}
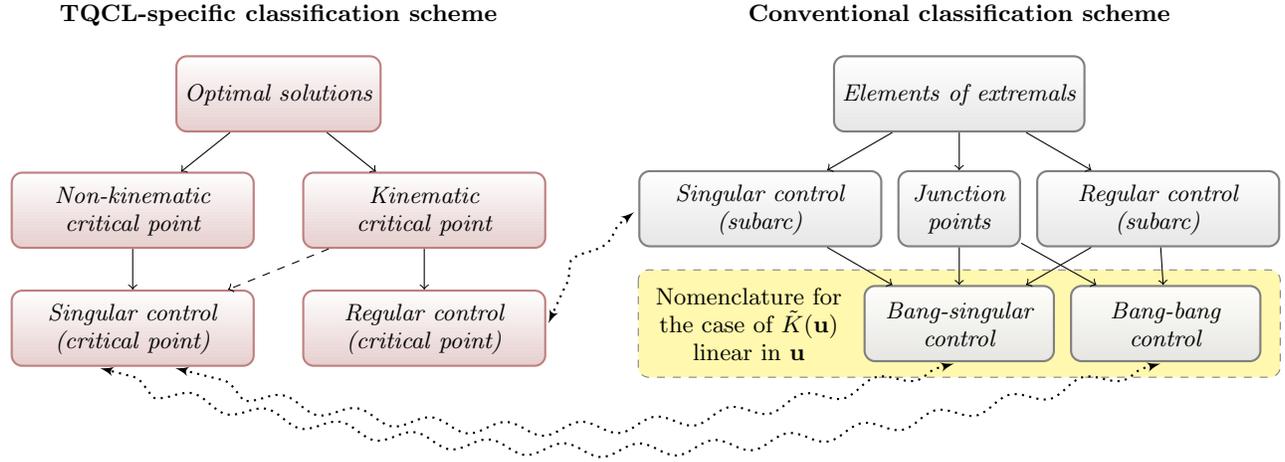
\begin{figure*}[bt]
\tikzstyle{format} = [ rectangle, 
                       rounded corners, 
                       thick,
                       minimum size=1cm,
                       draw=red!50!black!50,
                       top color=white,
                       bottom color=red!50!black!20,
                       font=\itshape]
\tikzstyle{format2} = [ rectangle, 
                       rounded corners, 
                       thick,
                       minimum size=1cm,
                       draw=black!50!black!50,
                       top color=white,
                       bottom color=white!50!black!20,
                       font=\itshape]

\begin{tikzpicture}[thick,
                     node distance=2cm,
                     text height=1.5ex,
                     text depth=.25ex]

\node[format2]  (Extremal) {Elements of extremals};
\node [format2,below=0.5cm of Extremal](JPoint)  {\parbox{1.4cm}{\centering Junction\\points}};
\node[format2,right=0.2cm of JPoint] (Regular) {\parbox{3cm}{\centering Regular control\\(subarc)}};
\node[format2,left=0.2cm of JPoint] (Singular) {\parbox{3cm}{\centering Singular control\\(subarc)}};

\node[format2,below=0.5cm of JPoint] (Bang-Sing) {\parbox{2.3cm}{\centering Bang-singular\\control}};
\node[format2,right=0.2cm of Bang-Sing] (Bang-Bang) {\parbox{2.3cm}{\centering Bang-bang\\control}};
\node[left=0.1cm of Bang-Sing] (Lin-Caution) {\parbox{2.6cm}{\centering Nomenclature for the case of $\tilde K(\vecu)$ linear in $\vecu$}};

    \begin{pgfonlayer}{background}
        \path (Singular.west |- Bang-Sing.north)+(-0.0,0.2) node (a) {};
        \path (Regular.east |- Bang-Bang.south)+(0.0,-0.2) node (b) {};
        \path [fill=yellow!40,rounded corners, draw=black!50, dashed] (a) rectangle (b); 
    \end{pgfonlayer}

\path[->] (Extremal) edge (Regular);
\path[->] (Extremal) edge (Singular);
\path[->] (Extremal) edge (JPoint);

\path[->] (Regular) edge (Bang-Bang);
\path[->] (Regular) edge (Bang-Sing);
\path[->] (JPoint) edge (Bang-Sing);
\path[->] (JPoint) edge (Bang-Bang);
\path[->] (Singular) edge (Bang-Sing);

\node[above=0.3cm of Extremal] (Caption){\textbf{Conventional classification scheme}};

\node[format,left=6cm of Extremal]  (Solutions) {Optimal solutions};
\node [below=0.8cm of Solutions](AuxNode)  {};
\node[format,right=0.2cm of AuxNode] (Kinematic) {\parbox{3cm}{\centering Kinematic\\critical point}};
\node[format,left=0.2cm of AuxNode] (Non-kinematic) {\parbox{3cm}{\centering Non-kinematic\\critical point}};
\node [below=1.1cm of AuxNode](AuxNode1)  {};
\node[format,right=0.2cm of AuxNode1] (Wrong-Regular) {\parbox{3cm}{\centering Regular control\\(critical point)}};
\node[format,left=0.2cm of AuxNode1] (Wrong-Singular) {\parbox{3cm}{\centering Singular control\\(critical point)}};

\path[->](Solutions) edge (Kinematic);
\path[->](Solutions) edge (Non-kinematic);
\path[->] (Kinematic) edge (Wrong-Regular);
\path[->,dashed] (Kinematic) edge (Wrong-Singular);
\path[->] (Non-kinematic) edge (Wrong-Singular);

\coordinate [right=0.5cm of Wrong-Singular.south](AuxCoord1)  {};
\coordinate [left=0.5cm of Wrong-Singular.south](AuxCoord2)  {};

\draw[<->, >=latex', shorten >=2pt, shorten <=2pt, bend right=-15, thick, dotted,decorate,decoration={snake,segment length=10mm,pre length=3mm,post length=3mm}] 
    (Bang-Bang.south) to (AuxCoord2); 
\draw[<->, >=latex', shorten >=2pt, shorten <=2pt, bend right=-15, thick, dotted,decorate,decoration={snake,segment length=10mm,pre length=3mm,post length=3mm}] 
    (Bang-Sing.south) to (AuxCoord1); 
\draw[<->, >=latex', shorten >=2pt, shorten <=2pt, bend right=15, thick, dotted,decorate,decoration={snake,segment length=10mm,pre length=3mm,post length=3mm}] 
    (Singular.west) to node[auto, swap] {}(Wrong-Regular.east); 

\node[above=0.3cm of Solutions] (Caption){\textbf{TQCL-specific classification scheme}};

\end{tikzpicture}
\caption{The standard and TQCL-specific classifications of optimal solutions. The waved dotted curves depict the approximate relationships between the classifications according to the conclusions of this paper.\label{*fig.01}}
\end{figure*}

Popularity of TQCL in the first decade of XXI century stimulated widespread usage of the corresponding field-specific scientific terminology. A part of TQCL terminology is incompatible with conventions used for decades by the rest of optimal control community. Currently, these two terminological systems coexists and often misleadingly associated to each other (see e.g. \cite{2011-Brif}, Sec. 1.7.1.1). The aim of this section is to clarify the differences between these two systems and summarize the conventions used in this paper.

The outline of two classifications of optimal solutions is sketched in Fig.~\ref{*fig.01}. The conventional classification on the right side is created primarily from engineering perspective and is based on distinguishing the different classes of subarcs and junction points of the extremals $\{\tilde\Psi(t),\tilde\vecu(t),\tilde\vecx(t)\}$. The specification of the nomenclature used in this scheme is detailed in the main text 
and Appendix~\ref{@SEC:APP-PMP}.

In contrast, the classification on the left-hand side is grounded on relating the technically reachable optimal solution to the limits established by laws of physics. Denote as $\Prop_{\tf,\ti}$ the propagator entering into the formal solution $\rho(\tf){=}\Prop_{\tf,\ti}\rho(\ti)$ 
of the master equation \MTeqref{01.01.2-Liouville_equation}:
\begin{equation}\label{01.01-evolution_superoperator}
\Prop_{\tf,\ti}(\vecu){=}\overrightarrow{\exp}\bigg(\int_{\ti}^{\tf}\Lvn(\vecu(t))dt\bigg).
\end{equation}
\noindent The reference solution called ``kinematic regular control'' (or ``kinematic regular critical point'' of the map $\vecu{\to}J$) represents the case when the optimal propagator $\tilde\Prop_{\tf,\ti}$ exactly coincides with the best quantum-mechanically admissible one (so-called ``kinematic limit''), and we have the total freedom in exploring the effect of all of the adjacent suboptimal propagators by introducing the small first-order variations to optimal controls $\tilde\vecu$ (i.e.~the differential map $\partial\tilde\vecu{\to}\partial\tilde\Prop_{\tf,\ti}$ is surjective). Other possible types of solutions are summarized in Table \ref{*TAB:TQCL-class}. 

\begin{table}[h!]
\caption{The classifcation of optimal solutions in TQCL.\label{*TAB:TQCL-class}}
\begin{center}
\newcolumntype{C}[1]{>{\centering\let\newline\\\arraybackslash\hspace{0pt}}m{#1}}
\renewcommand{\arraystretch}{1.5}
\begin{tabular}{@{}lC{2.5cm}C{2.5cm}}
\toprule
\multirow{2}{3cm}{\scriptsize Relative to the map $\Prop_{\tf,\ti}{\to}J$, the propagator $\tilde\Prop_{\tf,\ti}$ is:}
&\multicolumn{2}{c}{The map $\partial\tilde\vecu{\to}\partial\tilde\Prop_{\tf,\ti}$ is:}\\
\cmidrule[0.04em](r){2-3}
& surjective & rank deficient\\
\midrule
critical point
&{kinematic regular control}&{kinematic singular control}\\
regular point
&$\Cross$&{Non-kinematic singular control}\\
\bottomrule
\end{tabular}
\end{center}
\end{table}

The most confusing terminological clashes between the classification schemes are summarized in Table~\ref{*TAB:TQCL-clashes}.
\begin{table*}
\caption{Clashes between the regular terminology and TQCL-specific language.\label{*TAB:TQCL-clashes}}
\begin{center}
\newcolumntype{A}{@{\hskip8pt}m{7cm} @{\hskip8pt}}
\begin{tabular}{>{\textbf}l A A}
\toprule
 Term &  TQCL-specific meaning & Usual meaning\\
\midrule
&&\\[-8pt]
\textbf{regular control}&
{The map $\delta\vecu{\to}\delta\Prop_{\tf,\ti}$ is surjective at optimal solution $\vecu{=}\tilde\vecu$.
\newline Synonim: \emph{regular critical point}}&
{The point or segment of extremal $\{\tilde\Psi,\tilde\vecu,\tilde\vecx\}$ where the control value $\tilde\vecu$ can be directly deduced from PMP \eqref{APP_PMP-PMP}}
\\
&&\\[-8pt]
\textbf{singular control}&
{The map $\delta\vecu{\to}\delta\Prop_{\tf,\ti}$ is non-surjective (rank deficient) at optimal solution $\vecu{=}\tilde\vecu$.
\newline Synonim: \emph{singular critical point}}&
{The point or segment of extremal $\{\tilde\Psi,\tilde\vecu,\tilde\vecx\}$ where the control value $\tilde\vecu$ cannot be directly deduced from PMP \eqref{APP_PMP-PMP}}\\
&&\\[-8pt]
\bottomrule
\end{tabular}
\end{center}
\end{table*}
Figure~\ref{*fig.01} highlights the fact that some terms have nearly opposite meanings: ``regular control'' on the TQCL side in most of the cases is characterized as singular according to traditional definition and \emph{vice versa}. Specifically, any kinematic critical point always corresponds to singular extremal (i.e., all of the constraints on controls are inactive and do not prevent reaching the best performance) whereas the non-kinematic critical points, according to results of this work, are mostly represented by bang-singular and, in some cases, by bang-bang extremals.

In this paper, we everywhere follow the conventional classification of the right side of Fig.~\ref{*fig.01}.

\section{Sewing conditions at junction points of extremals of the problem (\ref{SOM-01.01-Problem_settling}) with a single control parameter \texorpdfstring{$u_1$}{Lg}\label{@SEC:APP-extremals}}
According to PMP, the following necessary condition should hold at the points of discontinuity of optimal control $\tilde u_1(t)$ or any of its time derivatives:
\begin{gather}
\tilde K_{u_1}(t){=}\matel{\psi(t)}{\sLvn_1}{\rho(t)}{=}0.\label{01.01.4-Special_points_constraints-1}
\end{gather}
Eq.~\eqref{01.01.4-Special_points_constraints-1} is the only sewing condition of equality type (other than continuation requirements \eqref{APP_PMP-sewing-conditions}) for connecting two regular subarcs. However, additional constraints might be necessary in other cases. 

Let us consider the sewing conditions at the left endpoint $t_1$ of a singular arc $t{\in}[t_1,t_2]$. 
Note that the equality \eqref{01.01.4-Special_points_constraints-1} holds everywhere in interior of singular arc by definition. Hence, the following set of equalities is valid for any $t{\in}[t_1,t_2]$:
\begin{gather}
\der{}{t}\tilde K_{u_1}(t){=}\matel{\tilde\psi(t)}{[\sLvn_1,\sLvn_0]}{\tilde\rho(t)}{=}0,\label{01.01.4-Special_points_constraints-2}\\
\der{^{n+2}}{t^{n+2}}\tilde K_{u_1}(t){=}\matel{\tilde\psi(t)}{[[\sLvn_1,\sLvn_0],\sLvn_1]}{\tilde\rho(t)}\der{^{n}\tilde u_1(t)}{t^{n}}{+}\notag\\f(\der{^{n{-}1}\tilde u_1(t)}{t^{n{-}1}},...,\tilde u_1(t)){=}0,~~(n{=}0,1,2,...)\label{01.01.4-PF-derivatives}
\end{gather}

Eqs.~\eqref{01.01.4-Special_points_constraints-1} and \eqref{01.01.4-Special_points_constraints-2} represent the two necessary conditions for $t{=}t_1$ to be a switching point from regular subarc at $t{<}t_1$. Suppose that the inequality
\begin{gather}
\matel{\tilde\psi(t)}{[[\sLvn_1,\sLvn_0],\sLvn_1]}{\tilde\rho(t)}{\ne}0\label{01.01.4-non-branch-point}.
\end{gather}
holds at $t{=}t_1$. Then, Eqs.~\eqref{01.01.4-PF-derivatives} allow to sequentially determine both the singular optimal control $\tilde u_1(t_1{+}0){=}\frac{\matel{\tilde\psi(t_1)}{[[\sLvn_1,\sLvn_0],\sLvn_0]}{\tilde\rho(t1}}{\matel{\tilde\psi(t_1}{[[\sLvn_1,\sLvn_0],\sLvn_1]}{\tilde\rho(t_1}}$ and all of it's Taylor expansion coefficients $\der{^{n}\tilde u_1(t)}{t^{n}}\big|_{t{=}t_1{+}0}$ at the beginning of singular arc. Hence, the optimal process $\tilde v(t){=}\{\tilde\Psi(t),\tilde\vecu(t),\tilde\vecx(t)\}$ can be uniquely reconstructed along the entire singular arc $t{\in}[t_1,t_2]$ from its known parameters $\tilde v(t_1{-}0)$ at the right endpoint $t{=}t_1{-}0$ of adjacent regular arc. 

What if the inequality \eqref{01.01.4-non-branch-point} is violated at $t{=}t_1$? We will conventionally call such junctions as \emph{branch points}. 
We will say the branch point $t{=}t_1$ has order $s$ if none of equations \eqref{01.01.4-PF-derivatives} with $n{<}s$ can be resolved relative to $\tilde\vecu(t_1{+}0)$ or any of its time derivatives. In this case, the reconstruction of $\tilde v(t{>}t_1)$ from $\tilde v(t_1{-}0)$ is generally ambiguous and might depend on up to $s{-}1$ undetermined continuous parameters. Branch points might be both the junctions of regular and singular as well as of two singular subarcs. The necessary conditions for the branch point of $s$-th order include in addition to Eqs.~\eqref{01.01.4-Special_points_constraints-1} and \eqref{01.01.4-Special_points_constraints-2} at least $\frac12(s{+}1)(s{+}2){-}1$ extra equalities
\begin{gather}\label{01.01.4-branch-point}
\der{^l}{\alpha^l}\matel{\psi(t)}{(\ad_{\sLvn_0{+}\alpha\sLvn_1}^m[\sLvn_1,\sLvn_0])}{\rho(t)}\bigg|_{\alpha{=}0}{=}0\\\notag(l{=}0,...,m,~m{=}1,...,s),
\end{gather}
where $\ad_{\sLvn}{\odot}=\sLvn{\odot}{-}{\odot}\sLvn$. It is worth to remember, however, that the constraints \eqref{01.01.4-Special_points_constraints-1} and \eqref{01.01.4-Special_points_constraints-2} are non-specific for junctions of two singular subarcs since they are satisfied everywhere in their interiors. 

\section{Proof of theorem~\texorpdfstring{\MTref{01.01-Lemma-1_substitute}}{}\label{@SEC:APP-lemma}}
The proof of the Statement~\MTref{01.01-Lemma-1_substitute} relies on the sewing conditions identified in Appendix~\ref{@SEC:APP-extremals}. From this analysis it follows that the extremal $\{\ket{\tilde\rho(t)},\bra{\tilde\psi(t)},\tilde u(t)\}$ is splitted into $N_{\idx{spec}}{+}1$ smooth regular and singular subarcs by $N_{\idx{spec}}{\geq}0$ special junction points $t{=}\tau_k$ $(k{=}1,...,N_{\idx{spec}})$ which may include corner points as well as $N_{\idx{branch}}$ branch points of various orders $s_k$. Furthermore, the entire extremal is reconstructible from known initial state and costate variables $\ket{\tilde\rho(\ti)}$, $\bra{\tilde\rho(\ti)}$, the localization of the special points and the additional parameters enabling the selection of proper branch at the branch points. This means that the extremal problem \eqref{SOM-01.01-Problem_settling} can be equivalently restated as the problem of determining $P_{\idx{total}}$ parameters which include $N_{\idx{spec}}$ times $\tau_k$, $2N^2$ elements of state and costate vectors $\ket{\tilde\rho(0)}$ and $\bra{\tilde\psi(0)}$, plus additional $P_{\idx{branch}}{\leq}\sum_{k{=}1}^{N_{\idx{branch}}}(s_k{-}1)$ parameters needed to define the next branch to jump on, and, possibly the control time $\Tctrl$. The total number $P_{\idx{total}}$ of unknowns is
\begin{gather}\label{APP_Lemma_1-P_total}
P_{\idx{total}}{=}N_{\idx{spec}}{+}2N^2{+}P_{\idx{branch}}{+}\alpha,
\end{gather}
where $\alpha{=}1$ in the case of unconstrained optimization time and 0 otherwise. These parameters are subject to $C_{\idx{total}}$ constraints of equality type which include $2N^2$ boundary conditions \eqref{SOM-03.01-DEF:terminal_control} or \eqref{SOM-03.01-DEF:periodic_control}, one extra equality \eqref{SOM-03.01-K=0(free_time)} in the case of optimization with unconstrained time, and the certain number of special constraints at junction points between various subarcs. Specifically, we need one extra constraint \eqref{01.01.4-Special_points_constraints-1} at any corner point between regular subarcs, two constraints of form \eqref{01.01.4-Special_points_constraints-1}, \eqref{01.01.4-Special_points_constraints-2} at any junction from regular to singular subarc or at $t{=}\ti$ if the first subarc is singular and at least $C_{\idx{branch}}{\geq}\sum_{k{=}1}^{N_{\idx{branch}}}(\frac12(s_k{+}1)(s_k{+}2){-}1)$ additional constraints \eqref{01.01.4-branch-point} identifying the branch points. Hence, the total number of constraints of equality type is
\begin{gather}\label{APP_Lemma_1-C_total}
C_{\idx{total}}{=}2N^2{+}\alpha{+}N_{\idx{spec}}{-}N_{\idx{sing}}{+}C_{\idx{branch}}{+}\beta,
\end{gather}
where $N_{\idx{sing}}$ is a number of branch points connecting singular arcs and $\beta{\in}[0,2]$ is the number of singular terminating arcs adjacent to $t{=}\ti$ and $t{=}\tf$. In absence of accidental redundancies the resolvability of the optimization problem requires that $P_{\idx{total}}{-}C_{\idx{total}}{\geq}0$, i.e.:
\begin{gather}\label{APP_Lemma_1-resolvability}
0{\leq}P_{\idx{branch}}{+}N_{\idx{sing}}{-}C_{\idx{branch}}{-}\beta{\leq}\notag\\
\sum_{k{=}1}^{N_{\idx{branch}}}(s_k{-}\frac12(s_k{+}1)(s_k{+}2){+}1{-}\beta{\leq}{-\beta}.
\end{gather}
Here we accounted for the facts that $N_{\idx{sing}}{\leq}N_{\idx{branch}}$ and that $s_k{-}\frac12(s_k{+}1)(s_k{+}2){+}1{<}0$ for any integer $s_k{\geq}1$. Note that the last inequality turns into equality only in the case $N_{\idx{branch}}{=}0$. Thus, the resolvability criterion $P_{\idx{total}}{-}C_{\idx{total}}{\geq}0$ generally holds (as an equality) only for extremals without branching points and terminated by regular subarcs at both ends $t{=}\ti$ and $t{=}\tf$ Q.E.D.

\section{Proof of theorem~\texorpdfstring{\MTref{05.01-periodic_control_theorem}}{}\label{@SEC:APP-periodic_control_theorem}}
To prove the first part of Statement~\MTref{05.01-periodic_control_theorem}, assume that $\tilde u(t\text{~mod~}\Tctrl)$ is the globally optimal periodic solution. Then, $\tilde u$ should also be the optimal solution for the set of the periodic problems periods $nT$ $(n{=}1,2,...)$. Indeed, $\tilde u$ is admissible for these problems and affords the same values of the performance index: $J_{n\Tctrl}{=}J_{\Tctrl}$ which reaches the absolute maximum by assumption. Given that the system is open, the absolute values of all but one engenvalues of the Liouville propagator $\tilde\sProp_{\Tctrl,0}$ are less than unity: $|\lambda_k|{<}1$ ($k{>}1$), and the last eigenvalue $\lambda_1$ corresponds to the eigenvectors $\bra{1}$ and $\ket{\tilde\rho(0)}$, where $\bra{1}$ is the vector 
representation of the identity observable \cite{REFINEMENT-notations}. Since $\lim_{n{\to}\infty}\lambda_k^n{=}0$, all the eigenvalues of the Liouville superoperator $\tilde\sProp_{\Tctrl,0}^n$ except one tend to zero as $n{\to}\infty$. From this relation and the normalization condition \eqref{03.01-psi_&_rho_normalization} if follows that
\begin{gather}\label{05.01-psi-limit}
\lim_{n{\to}\infty}\bra{\tilde\psi_{n\Tctrl}(n\Tctrl)}\tilde\sProp_{nT,0}^n{=}0,
\end{gather}
where $\bra{\tilde\psi_{n\Tctrl}(t)}$ denote the optimal costate for the periodic optimization with period $n\Tctrl$.
Substituting \eqref{05.01-psi-limit} into the boundary condition \eqref{SOM-03.01-DEF:periodic_control} one obtains that 
\begin{gather}
\lim_{n{\to}\infty}\bra{\tilde\psi_{n\Tctrl}(t)}{=}\bra{O_{\idx{n}}},
\end{gather}
which coincides with the boundary constraint \eqref{SOM-03.01-DEF:terminal_control} on the costate variables for the terminal control problem \eqref{SOM-01.01-Problem_settling} which proves the first part of Statement~\MTref{05.01-periodic_control_theorem}.

The converse statement follows from the fact that if  $\tilde u(t\text{~mod~}\Tctrl)$ is the optimal solution of the terminal problem with control time $n\Tctrl$ where $n{\to}\infty$ when the PMP requires that  
\begin{gather}
\forall t:\sum_{m{=}1}^{n}\tilde u_1(t)\tilde K_{u_1}(t{-}m\Tctrl){\to}\max,
\end{gather}
where we accounted for the periodicity or $\tilde u_1(t)$. Performing the summation and again utilizing the properties of eigenvalues of $\sProp_{\Tctrl,0}$ one gets:
\begin{gather}
\matel{\tilde\psi_{\Tctrl}(t)}{\sLvn_1}{\tilde{\rho(t)}}{=}0,
\end{gather}
where $\bra{\tilde\psi_{\Tctrl}(t)}$ matches the boundary constraints \eqref{SOM-03.01-DEF:periodic_control} on the costate variables in the periodic control problem \eqref{SOM-01.01-Problem_settling} with the period $\Tctrl$. The latter fact completes the proof.

\section{Proof of theorem~\texorpdfstring{\MTref{04.01-projective_control_theorem}}{}\label{@SEC:APP-projective_control_theorem}}
Suppose that there exist controls $\tilde u_k(t)$, $k{\in}\kappa$ which are singular at some $t{=}t_0$ (we assume that $t_0$ is not a junction point for the rest of controls). Let $t_1$ and $t_2$ be the closest to $t_0$ left and right junctions or endpoints of the optimal solution $\tilde\vecu(t)$. Then, $\sLvn_{\idx{c}}{=}\sLvn_0{+}\sum_{l{\not\in}\kappa}\tilde u_l(t)\sLvn_l$ is the time-independent part of the Liouvillian for $t{\in}(t_1,t_2)$ \cite{REFINEMENT-notations}. 

Assume that $\kappa{\ne}\emptyset$. Then, application of PMP together with normality condition \eqref{03.01-psi_&_rho_normalization} leads to the equalities:
\begin{equation}\label{APP_relax-dK/du_k}
\pder{\tilde K(t)}{\tilde u_k}{=}\scpr{\tilde\psi(t)}{\rho_k}{=}0\mbox{ for any }k{\in}\kappa\mbox{ and }t{\in}(t_1,t_2).
\end{equation}
Using Eqs.~\eqref{03.01-psi_&_rho_normalization} and the equalities $\matel{\tilde\psi(t_0)}{\sLvn_{k}}{\rho_k}{=}0$ ($k{\in}\kappa$) which can be obtained by differentiating Eqs.~\eqref{APP_relax-dK/du_k} with respect to time, one can obtain by induction the relations:
\begin{equation}\label{APP_relax-d(dK/du_k)_dt}
\left.\der{^n}{t^n}\pder{\tilde K}{\tilde u_k}\right|_{t{=}t_0}{=}({-}1)^n\matel{\tilde\psi(t_0)}{\sLvn_{\idx{c}}^{n}}{\rho_k}{=}0,
\end{equation}
valid for any $k{\in}\kappa$ and arbitrary $n$. Applying Eq.~\eqref{APP_relax-d(dK/du_k)_dt} to arbitrary function $f$ of operator arguments $\sLvn_{\idx{c}}$ and $\sLvn_{k}$ (${k{\in}\kappa}$) we have:
\begin{equation}\label{APP_relax-valid_for_any_f}
\matel{\tilde\psi(t_0)}{f(\sLvn_{\idx{c}},\sLvn_{k{\in}\kappa})}{\rho_k}{=}\sum_{n}\sum_{k{\in}\kappa}f_{n,k}\matel{\tilde\psi(t_0)}{\sLvn_{\idx{c}}^{n}}{\rho_k}{=}0.
\end{equation}
The latter is possible when $\mathspan\{\sLvn{\idx{c}},\sLvn_k\}$ can be decomposed into a direct sum of two orthogonal subspaces. This case, however, is of minimal physical interest since it just describes the situation of lack of controllability. Thereby, $\kappa{=}\emptyset$ and all the controls $u_k$ are regular and take only either minimal or maximal admissible values Q.E.D.

\section{Proof of theorem~\texorpdfstring{\MTref{04.01-mixed_control_theorem}}{}\label{@SEC:APP-mixed_control_theorem}}
Suppose that the optimal solution $\tilde u_k(t)$ contains singular subarc embracing the time interval $t{\in}(t_1,t_2)$. Then, according to \eqref{APP_PMP-PMP} the PF should satisfy the equation \cite{REFINEMENT-notations}:
\begin{equation}\label{APP_mixed-dK/du_k}
\pder{\tilde K(t)}{u_{k}}{=}\scpr{\tilde\psi(t)}{\rho_k}{=}0
\end{equation}
along this subarc. Consider an arbitrary variation $\delta u_k$ of the projective control on the singular subarc: $\delta u_k(t){\ne}0$ only for $t{\in}(t_1,t_2)$. 
Denote as $\mathbb{U}_{T,0}^{[\delta u]}$ and  $J^{[\delta u]}$ the corresponding perturbed propagator and respective performance index, so that $\ket{\rho^{[\delta u]}(t)}{=}\mathbb{U}_{t,0}^{[\delta u]}\ket{\rho(t_i)}$. Let $\mathcal{F}[\delta  u]{=}\scpr{\tilde\psi(T)}{\rho^{[\delta  u]}(T)}$. It follows from \eqref{SOM-03.01-DEF:terminal_control} that
\begin{equation}\label{APP_mixed-F-1_st_condition}
\mathcal{F}[\delta u]{=}\scpr{O}{\rho^{[\delta  u]}(T){-}\tilde\rho(T)}{=}J^{[\delta u]}{-}\tilde J.
\end{equation}
On the other hand, $\mathcal{F}[\delta u]$ can be expanded in series as:
\begin{equation}\label{APP_mixed-F-2_nd_condition}
\mathcal{F}[\delta u]{=}(c_{0}{-}1)\mathcal{F}[0]{+}\sum _{k\in \kappa }\int _{0}^{T}c_{t}\delta u_{k}(t)\scpr{\tilde\psi(t)}{\rho_{k}} dt,
\end{equation}
where $c_{\tau}[\delta u]{=}e^{{-}\int_{\tau}^{T}\sum_{k{\in}\kappa}\delta u_{k}(t)dt}$. From \eqref{APP_mixed-dK/du_k} and \eqref{03.01-psi_&_rho_normalization} it further follows that
$\mathcal{F}[\delta u]{=}(c_{0}{-}1)\mathcal{F}[0]{=}0$. Combining this result with \eqref{APP_mixed-F-1_st_condition}, we finally come to equality $J^{[\delta u]}{=}\tilde J$ Q.E.D.

\section{Specification of the \texorpdfstring{$\Lambda$}{Lg}-system\label{@SEC:APP-Lambda-system}}
\begin{figure}[ht!]
\includegraphics[width=0.25\textwidth]
{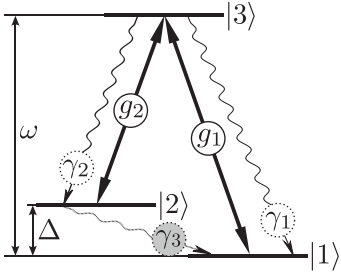}
\caption{The scheme of interactions in the periodically controlled $\Lambda$-system.\label{*fig.02} }
\end{figure}

In this Appendix the Dirac notations will be used in the conventional way for denoting the wavefunctions -- eigenstates of the eigen Hamiltonian of the quantum-mechanical system. The scheme of the periodically controlled $\Lambda$-system used in our example is sketched in Fig.~\ref{*fig.02}. The two ground levels $\ket{1}$ and $\ket{2}$ of this system are nearly degenerate having small energetic splitting $\Delta$ and directly dynamically coupled only by spontaneous transition $2{\leadsto}1$ with the decay rate $\gamma_3$, whereas the higher-lying third level is subject of both the coherent and spontaneous transitions: the former ones are induced by the control field $\vec{\cal E}$ and the latter ones lead to decay into one of the ground states with the equal rates $\gamma_1$ and $\gamma_2$.

This model is widely used in atomic physics to describe the nearly resonant interaction of atoms, quantum dots etc.~with optical or microwave radiation accounting for the fine or hyperfine structure of the atomic ground state or the effects of the Zeeman splitting of the ground state in the constant external magnetic field.

Since in our example the amplitude $|\vec A|$ of the frequency modulated laser field $\vec{\cal E}{=}\vec A\cos(\omega t{+}\int_{{-}\infty}^{t}u_T(t)dt)$ is assumed to be constant, the absolute values of the off-diagonal elements of the dipole-interaction Hamiltonian $\hat H_i{=}{-}\vec{\hat{\mu}}\vec{\cal E}$ are time-independent: $|\matel{1}{\hat H_i}{3}|{=}g_1$ $|\matel{2}{\hat H_i}{3}|{=}g_2$. The carrier frequency $\omega$ was chosen equal to the resonance transition frequency between the levels $\ket{1}$ and $\ket{3}$. All the three relaxation channels are assumed to be Markovian, and the corresponding Liouville superoperators ${\cal L}_{\idx{rel},i}$ are described within Linblad formalism:
$${\cal L}_{\idx{rel},i}\rho{=}\gamma_i\left({L}_{i}{\rho}{L}^{\dagger}_{i}{-}\frac{1}{2}({L}_{i}^{\dagger}{L}_{i}{\rho}{+}{\rho}{L}_{i}^{\dagger}{L}_{i})\right),$$
where ${L}_1{=}\ket{1}\bra{3}$, ${L}_2{=}\ket{2}\bra{3}$ and ${L}_3{=}\ket{1}\bra{2}$.

The actual values of parameters used in the calculations are summarized in the table:
\begin{center}
\newcolumntype{E}{@{\hskip4pt}c@{\hskip4pt}}
\begin{tabular}{>{\bf}m{2cm} E E E E E E }\toprule
\multirow{2}{*}
{Parameter}& $\Delta$  &   $g_1$    &   $g_2$    &   $\gamma_1$, $\gamma_2$   &    $\gamma_3$   \\
          &   [MHz]   &   [KHz]    &   [KHz]    &  [ms$^{-1}$]               &   [ms$^{-1}$]      \\\hline
Value     &    1.59   &    159     &    127     &   554                      &     236          \\\bottomrule
\end{tabular}
\end{center}

\end{document}